\documentclass[submission,copyright,creativecommons]{eptcs}
\providecommand{\event}{CASSTING'16} 
\usepackage{breakurl}             
\usepackage{epsf,graphicx}

\usepackage{color}
\usepackage{enumerate}
\usepackage{amsmath}
\usepackage{amsfonts}
\usepackage{amssymb}
\usepackage{latexsym}

\newcommand{\rmd}{\mathrm{d}}

\title{A Web-based Tool for Identifying Strategic\\ Intervention Points in Complex Systems\thanks{ This work was partly funded by the UK Research Council EPSRC, under the project {\it Evolution and Resilience of Industrial Ecosystems} (ERIE), Contract No. EP/H021779/1.}}
\author{Sotiris Moschoyiannis
\institute{Department of Computer Science}
\institute{University of Surrey, GU2 7XH, UK}
\email{s.moschoyiannis@surrey.ac.uk}
\and
Nicholas Elia 
\institute{Extended-Content Solutions}
\institute{London, SE1 9RT, UK}
\email{\quad nich.elia@gmail.com}
\and
Alexandra S. Penn
\institute{Department of Sociology}
\institute{University of Surrey, GU2 7XH, UK}
\email{a.penn@surrey.ac.uk}
\and
David J. B. Lloyd and Chris Knight
\institute{Department of Mathematics}
\institute{University of Surrey, GU2 7XH, UK}
\email{d.j.lloyd@surrey.ac.uk}
}

\def\titlerunning{A tool for exploring strategic intervention points}
\def\authorrunning{S. Moschoyiannis, N. Elia, A. S. Penn, D. J.B. Lloyd \& C. Knight}
\def\copyrightholders{S. Moschoyiannis, N. Elia,\\ A. S. Penn, D. J.B. Lloyd \& C. Knight}
\begin{document}
\maketitle

\begin{abstract}
Steering a complex system towards a desired outcome is a challenging task. The lack of clarity on the system's exact architecture and the often scarce scientific data upon which to base the operationalisation of the dynamic rules that underpin the interactions between participant entities are two contributing factors. We describe an analytical approach that builds on {\it Fuzzy Cognitive Mapping} (FCM) to address the latter and represent the system as a complex network. We apply results from network controllability to address the former and determine {\it minimal control configurations} - subsets of factors, or system {\it levers}, which comprise points for strategic intervention in steering the system. We have implemented the combination of these techniques in an analytical tool that runs in the browser, and generates {\it all} minimal control configurations of a complex network. We demonstrate our approach by reporting on our experience of working alongside industrial, local-government, and NGO stakeholders in the Humber region, UK. Our results are applied to the decision-making process involved in the transition of the region to a bio-based economy.
\end{abstract}

\section{Introduction}
\label{intro-sec}
We investigate a generic type of system that consists of a number of participating entities, including human and software agents acting for themselves or on behalf of organisations, which interact with one another and the environment. The interactions happen at multiple levels (micro and macro) and often reflect interdependencies between participating entities. The resulting {\it complex system} may evolve over time in often unpredictable ways. A range of systems in biology (brain, immune system, cell), nature (social insect, ant colonies), but also business and socio-technical systems, e.g., see \cite{Got07}, \cite{KRMM09}, \cite{May08}, have been approached using complexity theory \cite{Mi-K03}. 
We describe such a system in more detail, within the Humber region real world case study, as it might be instructive to look at an example of a real industrial ecosystem which can be understood in terms of a complex system. The study was carried out within the {\it Evolution and Resilience of Industrial Ecosystems} (ERIE) project\footnote{http://www.surrey.ac.uk/erie}, which is an interdisciplinary project aimed at combining computational tools and techniques with complexity science to produce decision-making frameworks for stakeholders and policy makers in industrial networks. 


{\bf Case Study: the Humber region, UK}. Within the context of ERIE, we have been working with the industrial ecosystem of the Humber region. We provide a brief outline here, more details can be found in \cite{PKL+13}. The Humber region comprises strategic energy generation facilities and infrastructure based around fossil fuels. The transition to a low-carbon economy is seen as an opportunity for the region. The required infrastructure and support industries largely exist already while feedstock is available from the substantial agricultural hinterland and bulk imports via the large local ports. The move is underlined by new investment in large-scale renewable energy technologies. A number of biodiesel and bio-ethanol facilities already exist and more are under construction. The region is expected to become the centre of a UK biofuel industry responsible for 50\% of UK production within the next 5 years. Significant investment is under way in energy from biomass and bio-waste facilities, alongside developments in bio-refinery to produce high-value chemicals. The estuary is also of national and international biodiversity and conservation importance, and due to climate change presents increasing flood risk management issues; both of these issues can cause friction over proposed development. Furthermore, neighbouring communities face significant socio-economic problems including unemployment and fuel poverty. 

It can be seen that development of the Humber region and its economy touches upon a number of {\it factors} which are entangled in a web of biophysical, industrial, economic, social, and governance systems, populated by many diverse actors or stakeholder groups. This makes automated analysis of the causal structure inherently challenging. Our motivation for gaining an understanding of the causal structure of this system is that various stakeholders involved want to steer or influence it. Typically, there are multiple objectives in such systems, e.g., one might want to increase certain aspects such as the bio-based energy production and jobs and, perhaps, decrease others such as flood risk.

The objective of our work is to produce computational techniques to aid in designing context-appropriate system interventions that can steer the system towards a chosen set of goals. This is a challenging problem and the reason is twofold.  Firstly, it is usually impractical - or at least expensive - to attempt to control or influence every factor within the system directly. Also, some factors are easier to control than others - factors such as 'international instability' and 'fossil fuel price' are largely influenced by factors outside the boundary of our focal system and therefore harder to control.  This motivates the investigation of control theory \cite{control-theory}, and more specifically structural controllability \cite{Lin74}, in an attempt to exploit the causal structure of the network of factors, without knowledge of the underlying dynamics. The idea is to identify a smaller subset that act as system {\it levers}, in the sense that a lever factor can be used to control numerous others; hence, the impact of one's input effort is maximised. 

Secondly, attempting to control or alter any one factor will influence other factors in the system and altering some factors will have more influence on the system than others. This points to the importance of identifying the connections between the different factors in a complex system. However, the lack of a specification of the system or scientific data leaves this aspect open to suggestion. Our experience with complex industrial networks such as that of the Humber region suggests that the decisions of the stakeholders play a key role in determining the outcome of the system's development. This means that it is paramount that the various stakeholders are involved in constructing a model of the system and setting up the links between factors. This motivates the investigation of participatory modelling and in particular, the application of {\it Fuzzy Cognitive Mapping} (FCM) \cite{FCM} in arriving at a weighted directed network that makes all the local knowledge of the system dynamics, and any underlying assumptions, explicit. 

In this paper, we describe an analytical tool that can be used to aid stakeholder groups in not only exploring the effects of certain courses of action but also in determining the strategic points of intervention in a complex network. 
Recent developments in network controllability \cite{LSB11} concerning the reworking of the problem of finding {\it minimal control configurations} allow the use of the polynomial time {\it Hopcroft-Karp} algorithm \cite{HoK73} instead of exponential time solutions. Drawing upon \cite{LSB11}, we adapt the algorithm to i) work with a directed graph rather than a bipartite graph, and ii) produce {\it all} minimal control configurations rather than merely find one, and iii) determine the factors in each such configuration rather than only calculate the size (cf Section \ref{sec:control-theory}). 
This underpins the analytical tool we have been developing for cooperative and strategic reasoning in complex systems (cf Section \ref{sec:step-by-step}). The so-called {CCTool}, for Complex Control Tool,  runs inside the browser and can be accessed at \url{http://cctool.herokuapp.com}.

The remainder of this paper is structured as follows. Section \ref{sec:partic-modelling} describes the use of FCM in a participatory modelling setting. Section \ref{sec:control-theory} outlines the application of structural controllability to networks. The combination of FCM and network controllability for strategic intervention is illustrated by means of our analytical tool in Section \ref{sec:step-by-step}. Section \ref{sec:concl} concludes the paper and outlines possible extensions of this work.

\section{Participatory modelling for adversarial and cooperative decision-making}
\label{sec:partic-modelling}

{\it Fuzzy Cognitive Mapping} (FCM) \cite{FCM} is widely used for problem-solving in situations where numerous interdependencies are thought to exist between the important components of a system, but quantitative, empirically-tested information about the forms of these interdependencies is unavailable \cite{SSK98}, \cite{SKC12}. The method aims to encapsulate the qualitative knowledge of expert participants or system stakeholders in order to rapidly construct a simple systems-dynamics model of a specified issue. It is particularly useful when decisions of stakeholders play an important role in determining the outcome of a system's development and in situations where stakeholder participation is desirable or even required \cite{OzO04}. 

The process of model construction consists of several stages: First, stakeholders generate and select key concepts/factors that are important influences on, or parts of, the system of interest. Factors can be from any domain (social, economic, physical, etc.) and may be qualitative or quantifiable. Second, causal influences between factors ('positive' (+) or 'negative' (-) links, depending on whether an increase in one factor causes an increase or decrease in the other) are discussed and decided on. 
Finally, participants rank and verbally describe the strengths of these influences between factors, ultimately producing a directed graph with weighted links, which we refer to as the {\it cognitive map} or FCM. 

\begin{figure}[ht]
\begin{center}
{\scalebox{0.7}{\includegraphics{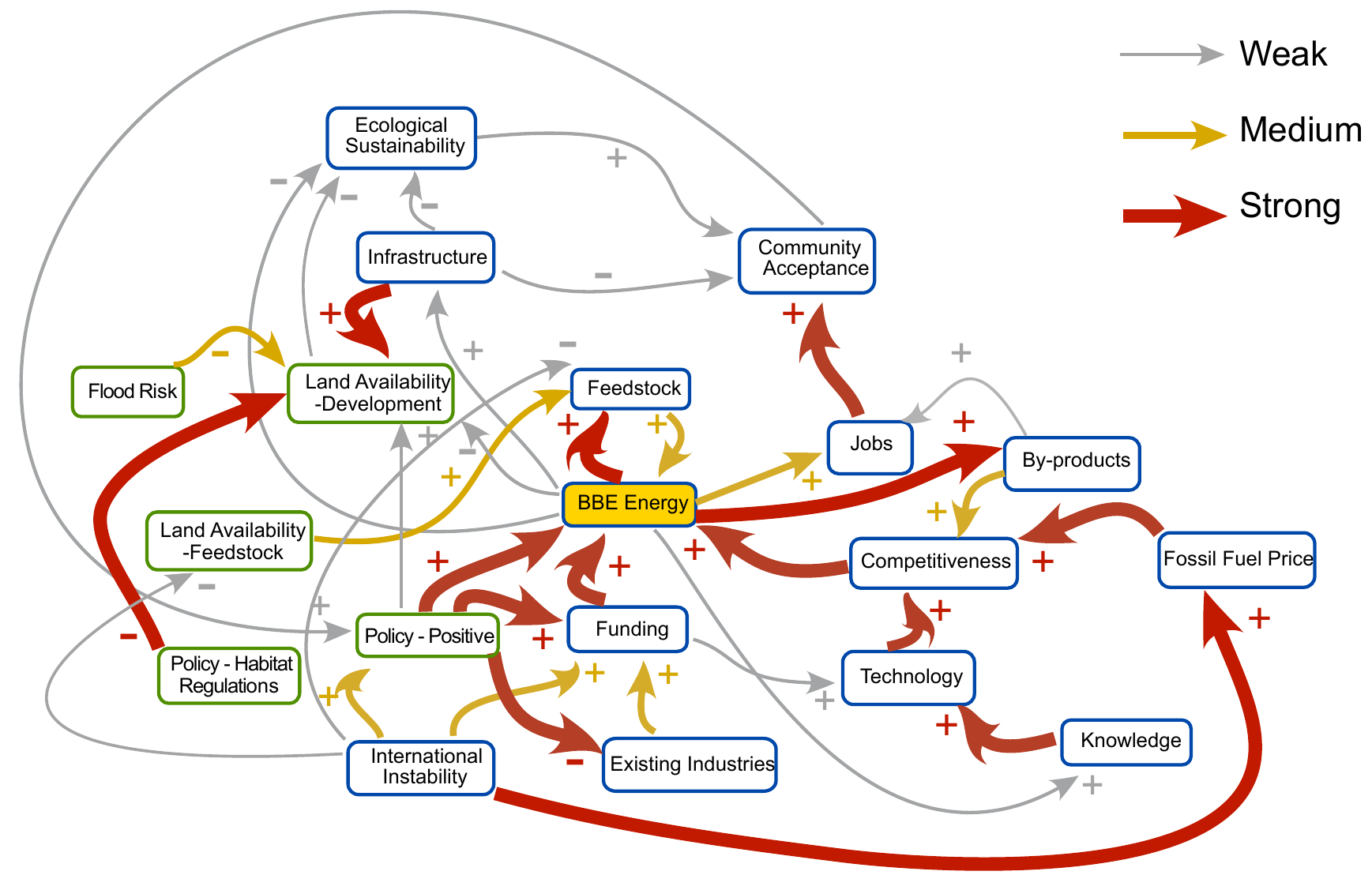}}}
\caption{FCM for the Humber region ecosystem - non-local feedstock production scenario}
\label{fig:SHB-FCM}
\end{center}
\end{figure}

As mentioned in the introduction, we are part of a team studying the Humber region in the UK. We held  a one-day participatory workshop with stakeholder groups (13 participants representing industry, local authorities, and NGOs) who collaboratively produced a single cognitive map, shown in Figure \ref{fig:SHB-FCM}. 

A verification and scenario generation session was held 3 months later at a local environmental managers' meeting. The participant group had similar composition with a few newcomers. Participants produced distinct, alternative structures for the cognitive map for two scenarios, namely  local and non-local feedstock production.  
The full details of the workshop methodology are found in \cite{SHBchapter} while the map output including transient dynamics (under both linear and sigmoidal mappings) are detailed in \cite{PKL+13}.

The FCM shown in Fig. \ref{fig:SHB-FCM} is a representation of the non-local feedstock supply scenario for the Humber region bio-based economy. In this scenario feedstock is imported via the local port rather than grown locally, meaning that there is no direct competition for land between feedstock production and industrial development. Availability of land is constrained by Habitat Regulations however. 

Factors shown in a rectangle with a green outline were added to the original FCM as a result of the verification exercise. International Instability (vs. UK stability), Flood Risk, and Habitat Regulations were identified as key external drivers of the Humber ecosystem. A {\it driver} is understood as a factor/node with outgoing links only. 
The thickness of links in Figure \ref{fig:SHB-FCM} denotes the strength of the influence (as done in \cite{PKL+13}).

The conversion of the FCM to an iterative map was done using two standard methodologies, creating a linear FCM and a sigmoidal FCM. With parameters chosen appropriately, as described in \cite{KLP14}, each of these iterative maps had a unique ordering of the factors at fixed points. In \cite{PKL+13} these orderings were used as a measure of importance via a form of sensitivity analysis. We compared the two orderings and if a factor was near the top of both rankings we said that that factor was one of the most important to the development of a bio-based energy economy in the Humber region. Similarly if a factor was near the bottom of both rankings then we described it as one of the factors of least importance (among the stakeholders' key concepts). This analysis revealed five factors consistently near the top of the rankings; Bio-based energy production (BBE Energy), By-products, Feedstock availability, Competitiveness and Jobs. It also revealed three factors as being of least importance; Land availability - development, Existing symbiotic industries and Ecological sustainability.

Simple dynamical models of fuzzy cognitive maps are commonly produced by transforming the weighted directed graph produced by stakeholders into an iterated map with the graph structure represented as an adjacency matrix and used to update a vector of arbitrary factor "values"
\[x_{n+1} = f(\mathbf{A}x_n)\]
where $x_0$ is given, $\mathbf{A}$ is the weighted connectivity matrix, $f$ is the thresholding function or functional mapping (which may take a variety of forms), and $n$ is the discrete time step. The state vector $x_n$ contains real values for all the key factors identified by participants. 

The outputs of such models are often used to "sense check" the graph structure, with the output used to rank factors in terms of relative importance. Despite the utility of dynamical models in aiding discussion, we found that different functional mappings gave inconsistent results when applied to the same map structure: often having more impact on model output than changes in the map itself (see \cite{PKL+13}, \cite{KLP14}). We thus decided to analyse our maps using graph theoretical methods. This means our  analysis relies only on the structural information provided by stakeholders and avoids the introduction of artifactual dynamics. Graph theory or network analysis has proved to be a useful tool in understanding whether specific network structures are vulnerable to failure and which particular nodes in a given network exert a strong influence on its processes, e.g., \cite{May08}. A network analytic approach (cf Section~\ref{sec:control-theory}) can be usefully applied to the causal inter-relations between factors produced in an FCM process as long as care is taken in the interpretation of results.


Further, we note that the approach to FCM described here differs to other approaches that use FCMs in non-participatory contexts (e.g., see \cite{Lin08}). The latter rely on exact functions used in the FCM process and seek to incrementally drive the system to a specific desired outcome. In a subjective setting however, where participating entities may represent physical, social, economic concepts (factors), it is impossible to represent the system in the level of accuracy that would be needed. Further, specifying exactly how a given factor should be controlled can be problematic in some cases (e.g., fossil fuel price). We argue, however, that it is another matter to determine {\it which} concepts should be controlled, i.e., what are the key points in the cognitive map where intervention can be strategic and most effective. The participatory methodology blended with FCM, as proposed here, enables the engagement of particular groups of stakeholders in determining which set of factors is the 'easiest' {\it for them} to influence in order to steer the complex system. This drives the investigation in the direction of control theory as applied to network analytics.

\section{Network controllability: control configurations of minimum size}
\label{sec:control-theory}

Control theory \cite{control-theory} largely advocates the ability to guide a system's behaviour towards a desired state, through a suitable choice of inputs. Previous work on applying control theory to communication networks  suggests that for most networks there exist various subsets of nodes (or factors, in our case) that can be used to control the whole network, in what is often referred to as {\it control configurations}. Trying to steer or control every possible node in a network is often infeasible and therefore it is useful to characterise these control configurations in a given network.  A network will typically have more than one control configuration and these will be of different sizes, i.e., containing a different number of nodes. Thus it seems appropriate to focus on {\it minimal control configurations}, i.e., those containing the least number of nodes. 

The minimal control node optimisation problem can be posed as an integer programming problem. We adopt the control notation from \cite{LSB11}, and assume the dynamics on the network can be linearised to yield the linear control problem
\begin{equation}\label{e:sys}
\frac{\rmd\mathbf{x}(t)}{\rmd t} = A\mathbf{x}(t) + B\mathbf{u}(t),
\end{equation}
where $\mathbf{x}(t)=(x_1(t), x_2(t),\ldots, x_N(t))^T$ is the vector of the state of the $N$ nodes of the network at time $t$, $A$ is the $N\times N$ weighted transposed adjacency matrix (where $a_{ij}$ is equal to the strength of the link (easy = 0.2, medium = 0.5, strong = 0.7) and zero otherwise), and $\mathbf{u}(t)=(u_1(t),u_2(t),\ldots,u_M(t))^T$ is the $M$ control inputs, and $B$ is the $N\times M$ ($M \leq N$) input matrix that identifies the nodes that are to be controlled (where the $j$-th column of $B$, $b_{ij}$ is equal to $\mathbf{u}_i$). 
We are interested in structural controllability where it is possible to choose non-zero weights of $A$ and $B$ such that the controllability matrix
\begin{equation}
C = (B,AB,A^2B,\ldots,A^{N-1}B),
\end{equation}
has full rank. We write this as $\mbox{rank}(\mbox{struct}(C))=N$. 
We may now define the minimal control node optimisation problem as the integer programming problem
\begin{equation}
\min_{\mbox{struct}(B)}(M),\qquad \mbox{subject to $\mbox{rank}(\mbox{struct}(C))=N$},
\end{equation}
where we wish to find the minimal number of control inputs required, over the structure of the set of input matrices $B$ (defined as struct({\it B})), subject to the system (\ref{e:sys}) being structurally controllable.


Liu {\it et al} in \cite{LSB11} investigate how the topology of a complex network affects its controllability. They show that there is a one-to-one mapping between control configurations of minimum size in a network and the {\it maximum matching} of the network. A maximum matching in a network refers to the maximum set of links that do not share start or end nodes. A node is said to be {\it matched} if a link in the maximum matching points at it. 
The importance of this development lies with turning a hard structural controllability problem to an equivalent geometrical problem on the network.  A network is {\it fully controlled} iff i) all unmatched nodes can be directly controlled\footnote{By 'directly controlled' we mean that each unmatched node is controlled by an independent time-varying input vector external to the network as defined previously. In practice, this refers to control exercised by some factor external to the system, e.g., stakeholder groups such as regional policy makers, industry actors, NGOs.}, and ii) there are directed paths from the inputs provided to all matched nodes \cite{YCK10}. 

Finding a maximum matching is a {\it colouring} problem on the network where the objective is to colour the maximum number of links across the network while respecting the constraint that there can be a maximum of one coloured link entering and a maximum of one coloured link leaving each node. Liu {\it et al} \cite{LSB11} showed that for a maximum matching the set of nodes which do not have a coloured link entering them form a control configuration of minimum size. There may be more than one such minimal control configurations. This reworking of the problem of finding a minimal control configuration allows to apply the well known {\it Hopcroft-Karp} algorithm \cite{HoK73}, which is a polynomial time algorithm instead of exponential time. Note that for sparse graphs (which one typically gets from FCMs) the Hopcroft-Karp algorithm has the best worst case complexity among other standard solutions. We exploit these developments in the algorithm applied to FCM maps of complex networks given as input to our analytical tool, discussed in the sequel. 

We have seen that Liu {\it et al.} in \cite{LSB11} showed that for a maximal matching the set of nodes (factors) which do not have a coloured edge entering them form a minimal control configuration. This reworking of the problem of finding a minimal control configuration as a colouring exercise on the network allowed the application of the well known polynomial time {\it Hopcroft-Karp} algorithm \cite{HoK73} in place of existing exponential time algorithms. The {\it Hopcroft-Karp} algorithm finds the size of the maximum matching in a given network and it does so by finding one such matching. According to the developments in \cite{LSB11}, this gives the size of the minimal control configurations of the network.

However, for our purpose of identifying 'levers' which can be used to exert strategic influence on the network towards a specific state, we need to examine {\it all possible} control configurations rather than one (which is sufficient for the purposes of \cite{LSB11} as it is adequate to give the size of a control configuration). In addition, and given the challenges of controlling a factor in practice as discussed in Section \ref{sec:partic-modelling}, we also want to know which nodes belong to each control configuration, although in the current work presented here we might have to make some concessions in this respect. The different combinations of factors to control comprise the potential alternative ways of steering the system towards a specific outcome.  

Finding all possible control configurations of minimum size is likely to be an NP-complete problem in the worst case scenario and hence the approach may not be applicable to larger networks. However, the work in \cite{DYphd} suggests otherwise as in its study of upland peat management it has considered networks comprising more than 200 nodes. FCMs tend to be sparse rather than dense networks. In any case, for larger networks we settle for getting an idea of how influential a factor or node is. This is given by how many control configurations each node appears in. In fact, finding that a node is in all configurations or none can be done in polynomial time. This is the first stage in the algorithm we have implemented for finding all possible control configurations of minimum size. The algorithm is outlined below, more details can be found in \cite{Kni13}.

\begin{enumerate}
\item Split graph into its connected components and treat each of these graphs separately, then
combine the control configurations for the separate graphs in all possible combinations.

\item  Run Hopcroft-Karp algorithm on graph of size $N$ to find the size of a maximum matching
($G_m$)

\item If $G_m = 0$ then there are $N$ control configurations, each consisting of one node. Return.

\item  Find every node which is never in a control configuration:
\begin{itemize} 
 \item For every node $i$:
   \begin{itemize}
     \item Create a subgraph S by removing all incoming edges to node $i$ in G.
     \item Run Hopcroft-Karp algorithm on S to find the size of a maximum matching ($S_m$).
     \item If $S_m \neq G_m$ then $i$ is never in a control configuration; otherwise, $i$ is in some
control configuration.
    \end{itemize}
 \item Return all nodes which are never in a control configuration $N_N$.
\end{itemize}

\item Find every node which is in all control configurations:
\begin{itemize}
 \item For every node $i \in N_N$. For every node $j$ which has an edge pointing to $i$ (in $G$):
  \begin{itemize}
   \item Create a subgraph $S$ by removing all outgoing edges from node $j$ and all incoming edges to node $i$ in $G$.
    \item Run Hopcroft-Karp algorithm on S to find the size of a maximum matching ($S_m$).
    \item If $S_m = G_m - 1$ then $i$ is not in all control configurations.
  \end{itemize}
 \item Return all nodes which are in all control configurations $N_A$. (Those where $S_m \neq
G_m - 1$ for all $j$ in the above.)
\end{itemize}

\item For every choice of $G_m - | N_A | (C_N)$ nodes from the nodes of $G - (N_A \cup N_N)$:
\begin{itemize} 
  \item Create a subgraph $S$ by removing all edges which point towards the nodes in $C_N$ in $G$. 
  \item Run Hopcroft-Karp algorithm on $S$ to find the size of a maximum matching ($S_m$).
  \item If $S_m = G_m$, then $C_N \cup N_A$ is a control configuration of minimum size.
\end{itemize}

\item Return list of all control configurations of minimum size. 
$\square$
\end{enumerate}


The Hopcroft-Karp algorithm is designed to run on bipartite graphs \cite{HoK73}. The network representation of the FCM for the complex system of the Humber region is a directed graph. To run the algorithm on this type of graph, we draw on \cite{LSB11} and make two copies of each node, one in the 'top' layer of the bipartite graph and one in the 'bottom' layer. Edges from the directed graph are then added with the outgoing node in the 'top' layer of the bipartite graph and the incoming node at the 'bottom' layer.
  
The worst case complexity of the Hopcroft-Karp algorithm is $N^{5 / 2}$ for dense graphs. The subsequent search of all the branches / matches is $2^N$. Hence, the  complexity of the algorithm is $2^NN^{5 / 2}$ to find the minimal control configurations. However, if the graph is too large\footnote{A network whose control configurations would take more than one minute to compute on a standard personal computer is understood as a 'large' network here. The different networks we have trialled it with in the Humber region take less than one second to compute. 
This amount of time allows the tool to be used effectively and seamlessly within a one-day participatory workshop to construct the input FCM.}
it quickly becomes too expensive to calculate the all possible control configurations. For this reason, the algorithm given above can be stopped before Step 6, in which case the complexity is $N^{9 / 2}$. This truncated polynomial time algorithm can be used to classify which nodes are in i) all, ii) none, or iii) some control configuration(s).  

Next, we describe how this computational technique has been coded up as the engine behind our analytical tool for complex systems and how this can used by stakeholders to design effective interventions.

\section{Using {\it CCTool} to identify points for strategic intervention}
\label{sec:step-by-step}

The main functionality of the tool is to take the FCM output of a participatory modelling workshop as its input, analyse the causal structure of the corresponding graph representation, and then generate and display all minimal control configurations to facilitate deliberations on the most effective courses of action. The Complex Control tool, {\it CCTool}\footnote{Available at \url{http://cctool.herokuapp.com}, please use {\bf username:} {\it CASSTING16} and {\bf password:} {\it demo}}, we have been developing is available as a web service for stakeholders, national and regional policy-makers, and researchers to use in cooperative decision-making in the process of steering complex systems. 
It is therefore important that the tool is i) easy-to-use with a professional 'look and feel', and ii) widely available from anywhere without the need for installation files on personal devices. 

\begin{figure}[ht]
\begin{center}
{\scalebox{0.4}{\includegraphics{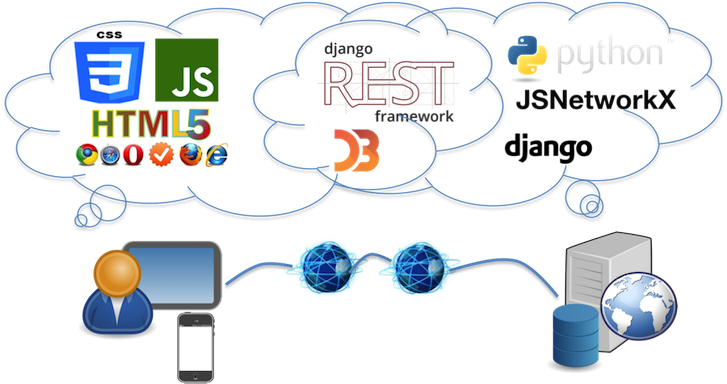}}}
\caption{Architectural view of the {\it CCTool} at \url{http://cctool.herokuapp.com}} 
\label{fig:arch}
\end{center}
\end{figure}

To achieve these objectives we have implemented {\it CCTool} as a responsive web tool. The prototype version demonstrated in this paper is built in Python using Django, a Python web framework for rapid development. The web presence is built using latest web technologies (HTML5, CSS3, JavaScript) which gives a fresh look and feel. To further enhance the user experience, we used the JSNetworkX (http://jsnetworkx.org) library which is based on the Python graph library NetworkX. This is largely responsible for the construction of the complex networks. In order to also provide a Web browsable API, we used the Django-Rest-Framework (http://www.djangoproject.com). The key aspects of the architecture of the tool as shown in Figure \ref{fig:arch}.

The tool supports all latest versions of modern web browsers and is accessible by any device which can run a browser (responsive design using the Bootstrap framework (http://getbootstrap.com)). 

\begin{figure}[ht]
\begin{center}
{\scalebox{0.3}{\includegraphics{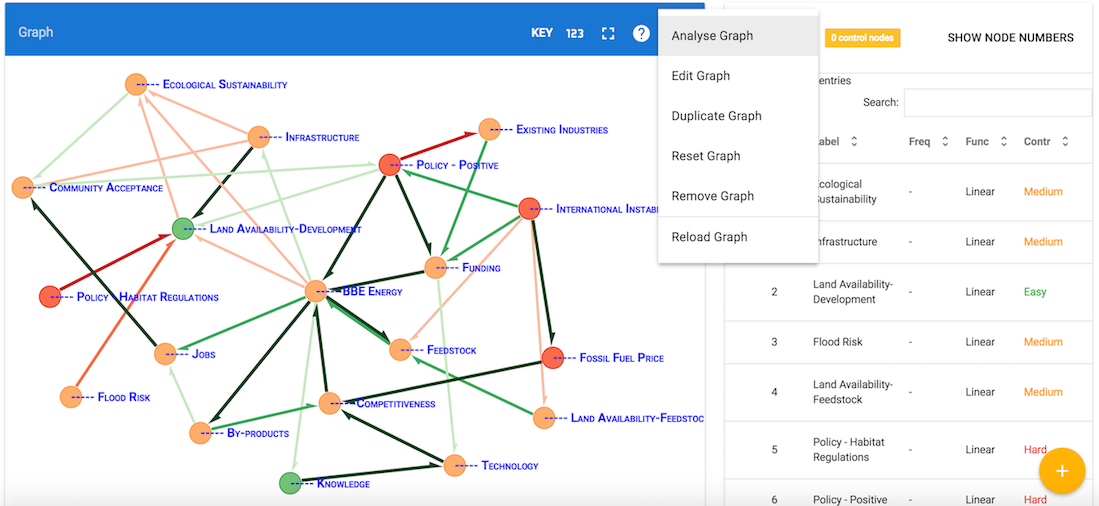}}}
\caption{The FCM represented as a directed weighted graph in the {\it ComplexControl} tool} 
\label{fig:BBE1graph}
\end{center}
\end{figure}

We now demonstrate how our analytical tool supports the exploration of the most effective points for intervention in a complex network. Figure \ref{fig:BBE1graph} shows the graph representation of the FCM for the Humber region we saw in Figure \ref{fig:SHB-FCM} (from Section \ref{sec:partic-modelling}). Such a graph can be added using the '+' icon at the bottom right of the {\it CCTool} window. 
 
.

The colouring on the nodes follows the traffic light scheme in denoting the controllability of the corresponding factor - green for easy, orange for medium, and red for hard to control. This can be set during the workshop using the menu on the right pane as shown in Figure \ref{fig:BBE1graph}. A similar scheme is applied to the links between nodes, with green denoting positive, red denoting negative and grey for neutral. The thickness of the link denotes the strength of the influence (weak, medium, strong). The drop down menu at the top right of the graph pane lists the actions available to the user (analyse, edit, duplicate, etc.).

\begin{figure}[ht]
\begin{center}
{\scalebox{0.3}{\includegraphics{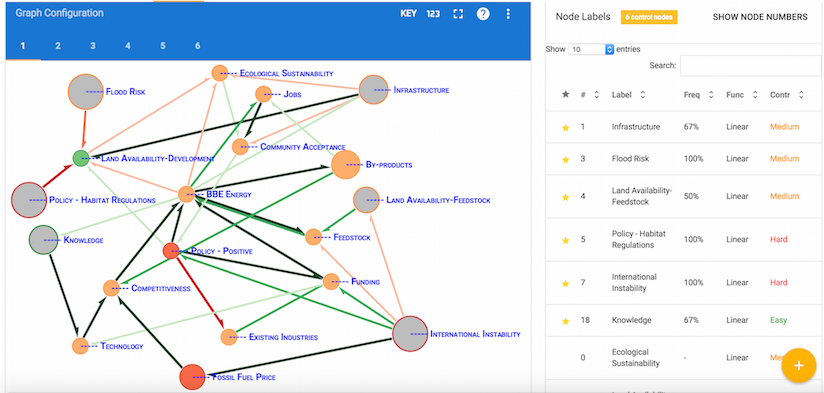}}}
\caption{A control configuration for the graph of Fig. \ref{fig:BBE1graph}}
\label{fig:BBE1control-config}
\end{center}
\end{figure}

A control configuration for the graph of Figure \ref{fig:BBE1graph}, corresponding to the non-local feedstock supply scenario for the Humber region in our case study is generated by applying the technique described in Section \ref{sec:control-theory}. The result is shown in Figure \ref{fig:BBE1control-config}. Nodes drawn in grey colour belong to the minimal control configuration shown here. The colour of the outline here denotes the ease of controllability of the corresponding node (green - easy, orange - medium, red - hard).

Hence, the factors Policy - Habitat Regulations, International Instability,  Infrastructure, Flood Risk, Land availability - feedstock, and Knowledge have been identified as control nodes within this complex network of factors at play in the industrial ecosystem of the Humber region.

Note that the size of a node in Figure \ref{fig:BBE1control-config} denotes how frequently that node appears in a control configuration, e.g., Flood risk here appears in more control configurations than Land availability - feedstock.  

It is also worth pointing out that there are more control configurations as indicated by the numbers 1-6) appearing in the blue ribbon above the control configuration in Fig. \ref{fig:BBE1control-config}. In fact, the Humber region complex system (for the non-local feedstock scenario, shown in Figure \ref{fig:BBE1graph}) has six distinct minimal control configurations. The controllability of each control factor (grey node) is used in our tool  to produce a total score for each configuration.  Compiling the results of the stakeholders' estimation of controllability, we ranked the control configurations (1-6) based on 'ease of control'. Due to space limitations, we only include the most controllable configuration here, as produced by the tool, shown in Figure \ref{fig:BBE1control-config}.

\begin{figure}[ht]
\begin{center}
{\scalebox{0.4}{\includegraphics{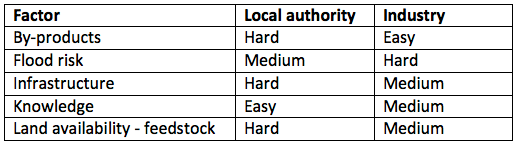}}}
\caption{Stakeholders' differing opinions on the controllability of certain factors}
\label{fig:diff-controllability}
\end{center}
\end{figure}

The ease of controllability (weak, medium, strong) assigned to any particular factor will inevitably vary between stakeholder groups as different types of actors can exert different types and extents of influence. In our particular case study of the Humber region (Section \ref{sec:partic-modelling}), it is instructive to consider the difference in controllability scores assigned by different groups of stakeholders, namely industrialists and local authorities, as shown in Figure \ref{fig:diff-controllability}.

\begin{figure}[ht]
\begin{center}
{\scalebox{0.32}{\includegraphics{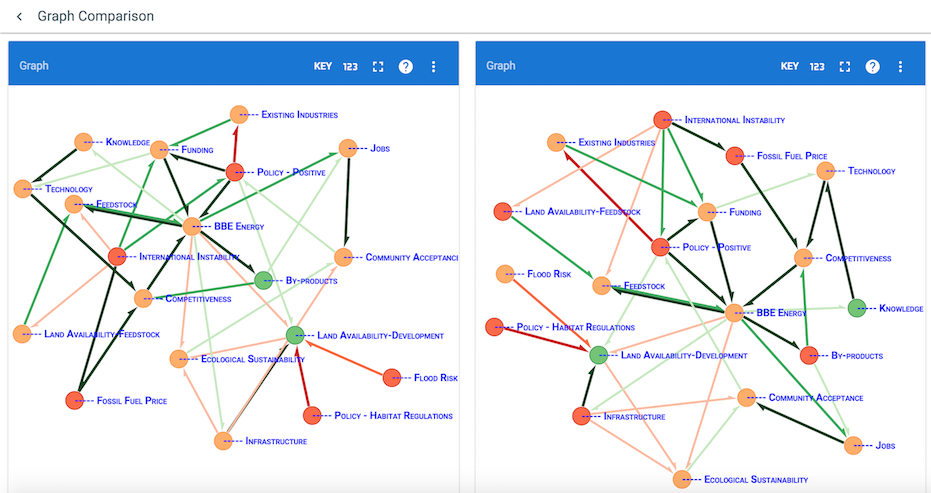}}}
\caption{Comparing the resulting graph structures; Local authority (left), Industry (right)}
\label{fig:compare-graphs-LA-Ind}
\end{center}
\end{figure}

If rather than using average controllability scores we examine the differences between different actors' perspectives we can gain useful information on for example, potential collaborators who might need to be brought on-board for a control strategy to be effective. Highlighting such differences between actors in a workshop context also serves as a means to increase understanding between different groups of their diverse perspectives and possibilities for action.  The effect on the corresponding graphs is shown in Figure \ref{fig:compare-graphs-LA-Ind}. 

\begin{figure}[ht]
\begin{center}
{\scalebox{0.3}{\includegraphics{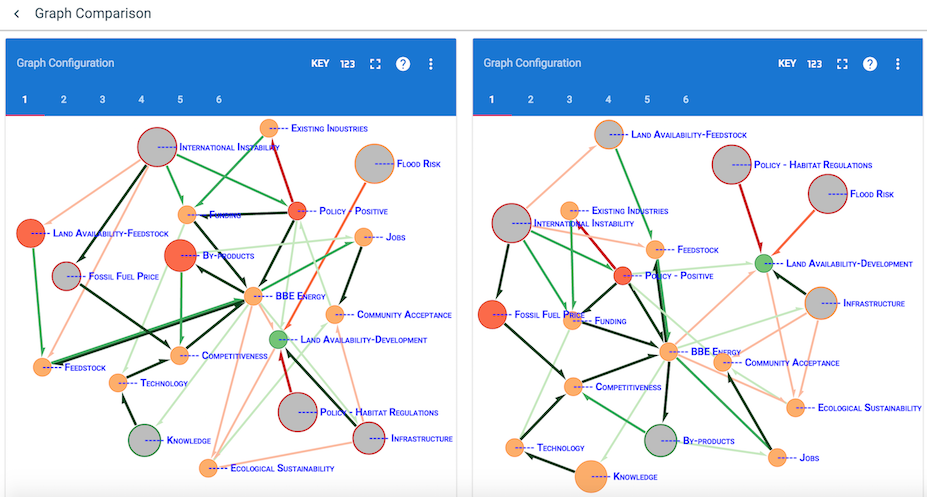}}}
\caption{Comparing the resulting control configurations; Local authority (left), Industry (right)}
\label{fig:compare-control-configs-LA-Ind}
\end{center}
\end{figure}

Views diverged on a number of factors, including some potential control nodes, meaning that the controllability ranking of control configurations would differ according to different stakeholder perspectives. Figure \ref{fig:compare-control-configs-LA-Ind} shows the control configurations of the respective graphs in Figure \ref{fig:compare-graphs-LA-Ind}.

As By-products and Infrastructure are both labelled as hard to control by local authorities, analysing the graph from this perspective suggests that all the most controllable configurations contain only one easy to control factor, Knowledge, and one with medium controllability, Flood risk. All other possible control nodes, Fossil fuel price, International instability, Habitat regulations and Infrastructure are considered as hard to control. This could evidently leave these stakeholders feeling that they have few options. From the Industry stakeholders' perspective on the other hand, By-products are considered easy and Infrastructure medium to control and these two factors form the backbone of their optimal control configurations. Combining, or even communicating, the two approaches, with local authorities influence over Flood risk and Knowledge, and Industry's perspective on Land availability and By-product manipulation leads to a broadening of the feasible options and opportunity for more effective intervention.

\begin{figure}[ht]
\begin{center}
{\scalebox{0.51}{\includegraphics{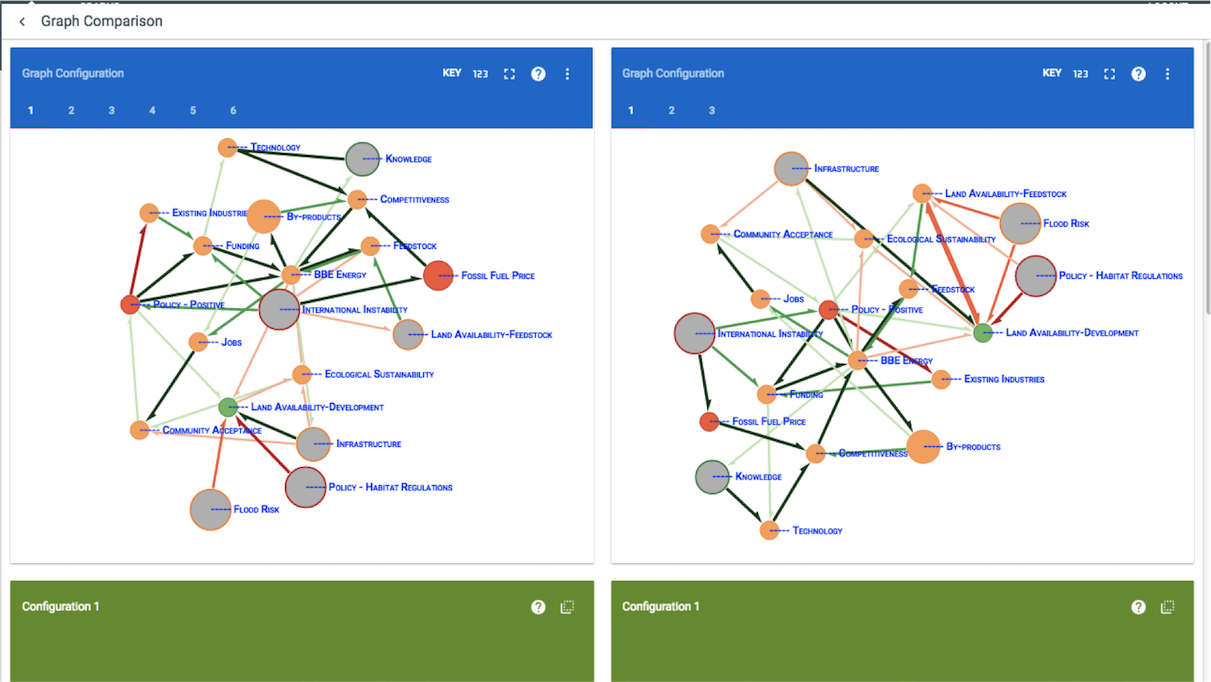}}}
\caption{Directed graphs for the FCM of the 2 scenarios: non-local (left) and local feedstock (right).}
\label{fig:diff-scenarios}
\end{center}
\end{figure}

The {\it CCTool} can also be used to compare different possible system scenarios represented as different graph structures. For example, in our participatory modelling workshops stakeholders produced the original graph in which feedstock for the Humber bio-based economy was assumed to be imported via the local ports. However, additional alternative scenarios were generated in follow-up workshops and discussion, including a locally-produced feedstock scenario. In this possible situation (shown in Figure \ref{fig:diff-scenarios}), Land availability for feedstock is subject to the same local influences as Land availability for development, that is, Habitat regulations and Flood risk. There also exists a negative interaction between Land availabilities for feedstock and development, as we assume that a finite amount of land is available.

On calculating the control configurations for these two different scenarios, we can immediately see a large difference. Land availability for feedstock is no longer a possible control node in the second scenario and the minimum control configurations for local feedstock scenario comprise sets of 5 control nodes rather than 6. There are also fewer possible configurations, only 3 rather than 6, meaning that there are fewer possible strategies that could be employed to steer the whole network. The most controllable configuration  in the local feedstock scenario contains Knowledge, Infrastructure and Flood risk as well as the hard to control Habitat regulations and International instability. Whereas the most controllable configurations in the imported feedstock scenario contain Land availability for feedstock in addition, with either By-products or Infrastructure as possible local control nodes (the one shown in Figure \ref{fig:diff-scenarios} contains Infrastructure). The reduction in number of options for full and effective control is interestingly accompanied by a need for fewer sources of control. Such differences can be used as a basis for discussion on the advantages and disadvantages of different scenarios from the perspective of system intervention.

\section{Concluding remarks and future work}
\label{sec:concl}

The representation of sets of interactions or relationships between interacting entities as a network or graph has become widespread in numerous fields \cite{BMBL09,PPP05}. Network analysis has proved to be a useful tool in understanding whether specific network structures are vulnerable to failure and which particular nodes in a given network exert a strong influence on its processes \cite{RMK09}. In this paper we have shown how a network analytic approach can be applied to the causal interrelations between factors produced in an FCM process as long as care is taken in the interpretation of results. For example, the findings in \cite{LSB11} suggest that driver factors in complex networks tend to avoid hubs.

Preliminary analysis on the structure of the weighted, directed graphs suggests that the method for finding all possible minimal control configurations is sensitive to certain patterns in the graph such as {\it self-loops} or {\it dilations} - the latter are also discussed in \cite{LSB11sup}. One way to identify such cases in an automated manner is to check whether the controllability matrix has a full rank. A non-full rank implies that minimal control configurations break down to smaller ones; 
effectively, different parts of the network ({\it stems}) spring out of each control node. It would appear that determining such stems is still useful as indication of what part of the network can be controlled from each node. This is a direction that deserves further investigation.  

Currently, our analytical tool produces the minimal control configurations for graphs without such features, each suggesting a way in which the system can be steered to exhibit some form of desired behaviour. We have based the analysis of the input FCM on its causal structure rather than functional mappings and combined this approach with results from network controllability. We have implemented this analytical technique in a web tool that can automatically identify subsets of nodes that can be used as {\it levers} to steer the whole system. The idea is that the information produced by our tool can be used imminently by stakeholders, during the participatory workshop, in deciding which course of action to take, e.g., based on how easy the factors in question are to influence or, more generally, how effective an intervention might be. 

The ranking between different minimal configurations is done in the current version of {\it CCTool} by considering the characterisation of the factors in terms of ease of control (easy, medium, hard). In some settings the interest might be in whether the system is vulnerable to external shocks, in which case we would be asking the expert participants to characterise or rank the factors based on how susceptible they are to shock effects. In other settings the interest might be in the (speed of) spread of a disease, so factors might be ranked by experts in terms of how likely they are to transmit to other factors in the network. 

The weights are currently only used in the analysis of teh FCM but not in determining the associated minimum control configurations, i.e., we work with a directed, not a weighted graph. There is leverage in incorporating different, perhaps more sophisticated, and customised mechanisms to accommodate other criteria for factors laid out by stakeholders in different problem domains. 


One limitation of the proposed approach is that it is not applicable when the factors have internal dynamics, i.e., when the nodes in the graph have self-loops. Preliminary analysis indicates that this is a challenging direction for future investigation. Another limitation has to with the sensitivity of the output to changes in the causal structure of the FCM. It would be interesting to investigate the effect of taking a node out of the network, by deleting the corresponding link. This would impact on the rankings and could possibly alter the control structure of the system. The extent of the effect, possibly in relation to the characteristics of the removed node, as well as the interplay with certain characteristics of the network (connectivity, sparsity, degree distribution) pose some challenging questions in future extensions of this work and versions of our analytical tool. 

Another limitation perhaps is that our approach will identify the most influential points for steering the system to a desired state, but it will not prescribe how to control or influence a factor. 
Although International Instability was indeed seen as impossible to control in the Humber case study, the other two drivers, Habitat Regulations and Flood Risk, were revealed to be somewhat controllable. In fact, 
the reasoning on imperfect information afforded by our computational tool in this instance revealed that the Flood Risk driver could be indirectly controlled via Land management or Infrastructure factors.

The aim of this analytical work is to provide a thinking tool for stakeholders in a given system so they can start to explore possibilities for effective management or policy interventions. The outputs of this control analysis on stakeholders' cognitive maps of their own systems can be used as the beginning of an intervention design process, involving further fact-finding and possibly empirical research and more detailed modelling. The control configurations provide a focus for such research. Although this work is at preliminary stages there are numerous ways in which stakeholders can use control information and the map construction functionality of {\it CCTool} to gain insight into their system.


\end{document}